\documentclass[12pt]{article}
\usepackage{graphicx}

\newcommand\pubnumber{NuPhys2016-DiDomizio}
\newcommand\pubdate{\today}

\def\unige{Dipartimento di Fisica\\
Universit\`a degli Studi di Genova, I-16146 Genova, ITALY}
\def\infnge{Istituto Nazionale di Fisica Nucleare\\
Sezione di Genova, I-16146 Genova, ITALY}

\def\Title#1{\begin{center} {\Large #1 } \end{center}}
\def\Author#1{\begin{center}{ \sc #1} \end{center}}
\def\Address#1{\begin{center}{ \it #1} \end{center}}

\newcommand\pubblock{\rightline{\begin{tabular}{l} \pubnumber\\
         \pubdate  \end{tabular}}}
\newenvironment{Abstract}{\begin{quotation}  }{\end{quotation}}
\newenvironment{Presented}{\begin{quotation} \begin{center} 
             PRESENTED AT\end{center}\bigskip 
      \begin{center}\begin{large}}{\end{large}\end{center} \end{quotation}}

\def\beq{\begin{equation}}
\def\eeq#1{\label{#1}\end{equation}}
\def\eeqn{\end{equation}}

\def\beqa{\begin{eqnarray}}
\def\eeqa#1{\label{#1}\end{eqnarray}}
\def\eeqan{\end{eqnarray}}

\let\bar=\overbar

\def\Dslash{\not{\hbox{\kern-4pt $D$}}}
\def\dslash{\not{\hbox{\kern-2pt $\del$}}}

\def\msb{{\bar{\ssstyle M \kern -1pt S}}}

\begin{document}
\begin{titlepage}
\pubblock

\vfill
\Title{Future prospects for neutrinoless double-beta decay}
\vfill
\Author{Sergio Di Domizio for the CUORE collaboration}
\Address{\unige}
\Address{\infnge}
\vfill
\begin{Abstract}
  The study of neutrinoless double-beta decay plays a fundamental role in the understanding of neutrino physics.
  Its observation would prove that neutrinos are Majorana particles and that lepton number is not conserved.
  Experimental searches demand detectors with a very large source mass and extremely low background.
  We report on planned future experimental searches and discuss their expected sensitivities.
\end{Abstract}
\vfill
\begin{Presented}
NuPhys2016, Prospects in Neutrino Physics\\
Barbican Centre, London, UK,  December 12--14, 2016
\end{Presented}
\vfill
\end{titlepage}
\def\thefootnote{\fnsymbol{footnote}}
\setcounter{footnote}{0}

\section{Introduction}
\begin{figure}[!b]
\centering
\includegraphics[width=0.65\textwidth]{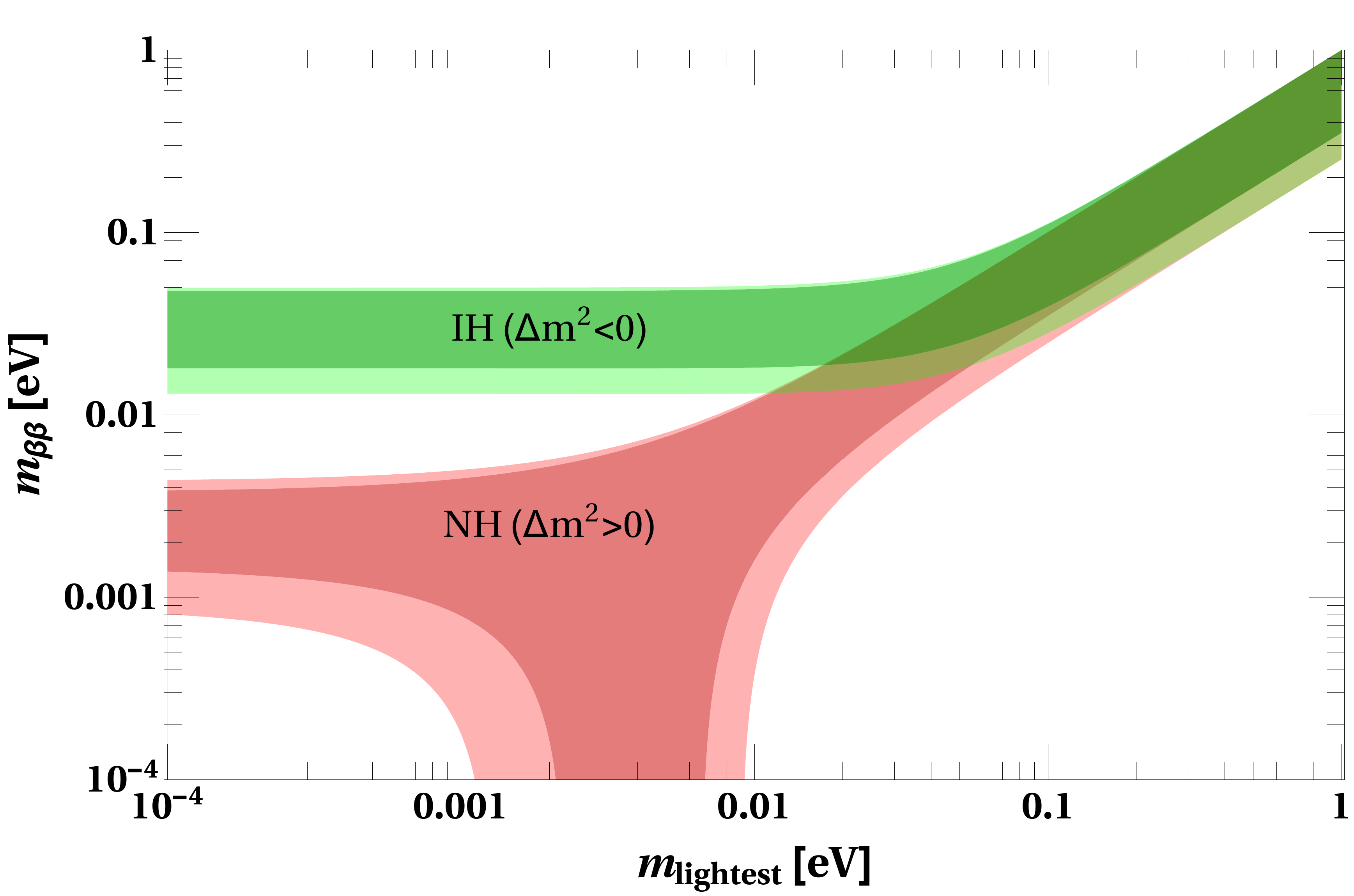}
\caption{Allowed value for $\left<m_{\beta\beta}\right>$ as a function of the mass of the lightest neutrino for normal (NH, red band) and inverted (IH, green band) neutrino mass hierarchy.
  Reprinted from~\cite{delloro}.}
\label{fig:mbb_mlight}
\end{figure}

Double-beta decay~\cite{DBDReview} is a rare spontaneous process in which the atomic number of a nucleus changes by two units.
It can only occur in some even-even nuclei where single-beta decay is energetically forbidden.
Two-neutrino double-beta decay mode (2$\nu$DBD) conserves the lepton number and is allowed by the Standard Model of particle physics.
It is the rarest decay ever observed, with half-lives in the range (10$^{19}$ -- 10$^{24}$)$\,$y~\cite{2nudbd}.
Neutrinoless double-beta decay (0$\nu$DBD) has never been observed and the half-life lower limits are in the range (10$^{21}$ -- 10$^{26}$)$\,$y.
Its observation would prove that the total lepton number is not conserved, and that neutrinos are Majorana particles~\cite{Valle}.
Under the assumption that 0$\nu$DBD takes place by the exchange of a light Majorana neutrino, the inverse of the decay half-life can be written as
\begin{equation}\label{eq:decayrate}
\frac{1}{T^{0\nu}_{1/2}} ~ = ~
G^{0\nu} \left| M^{0\nu}\right|^2 \frac{\left< m_{\beta\beta}\right>^2}{m_e^2},
\end{equation}
where $G^{0\nu}$ represents the decay phase-space, $M^{0\nu}$ is a matrix element that accounts for the nuclear part of the decay, $m_e$ is the electron mass, and $\left<m_{\beta\beta}\right>=\left|\sum_i{U^2_{ei}m_i}\right|$ is a function of the neutrino masses and the neutrino mixing matrix elements.
Formula~\ref{eq:decayrate} links the experimentally measurable decay half-life to $\left<m_{\beta\beta}\right>$, which encloses information about neutrino properties.
$\left<m_{\beta\beta}\right>$ can be expressed as a function of four unknown parameters: the sign of $\Delta$m$^2_{23}$, the mass of the lightest neutrino, and two Majorana phases.
This is represented in fig.~\ref{fig:mbb_mlight}, reprinted from~\cite{delloro}.

If the assumption of the exchange of a light Majorana neutrino is valid, then 
0$\nu$DBD could also give information on the neutrino mass hierarchy and the absolute scale.
Moreover this assumption allows to compare the sensitivity of 0$\nu$DBD searches based on different isotopes.
This comes at the cost of introducing uncertainties from the theoretical calculation of $M^{0\nu}$, which cannot be performed exactly.
Several models exist that make different approximations, leading to results whose reliability is difficult to assess.
A big effort has been put in these calculations in recent years, resulting into a better agreement between different models.
See e.g.~\cite{Engel} for a recent review.

In principle 0$\nu$DBD has a clear experimental signature.
The sum-energy of the two emitted electrons is fixed and equal to Q$_{\beta\beta}$, the Q-value of the decay.
This signal is qualitatively different from that of 2$\nu$DBD, in which part of the energy is carried away undetected by the two anti-neutrinos, resulting in a continuous energy spectrum extending up to Q$_{\beta\beta}$.
The value of Q$_{\beta\beta}$ and other relevant quantities are reported in table~\ref{tab:isotopes} for a selection of experimentally interesting isotopes.
Q$_{\beta\beta}$ lies in the energy range of natural radioactivity, which represents the dominant source of background for experiments.
Since the signal must be maximized and the background minimized, all experiments searching for 0$\nu$DBD share the need for a large number of source isotopes, a low background and a good energy resolution.

After a brief review of the current status of the experimental searches, we consider as a target sensitivity for future experiments the full exploration of the values of $\left<m_{\beta\beta}\right>$ corresponding to the inverted-hierarchy region of the neutrino mass.
We then conclude discussing the experimental techniques that will be used and some of the proposed future experiments.

\begin{table}[bt]
\begin{center}
\begin{tabular}{r|cccc}
  Isotope &  Q$_{\beta\beta}$ [keV] & i.a. [\%] & T$_{1/2}$(0$\nu$DBD) [10$^{25}\,$y] & T$_{1/2}$(2$\nu$DBD) [10$^{21}\,$y] \\
  \hline
  $^{76}$Ge & 2039 & 7.61 & $>$5.3~\cite{Ge0nu} & 1.65$^{+0.14}_{-0.12}$\\
  $^{82}$Se & 2995 & 8.73 & $>$0.036~\cite{Nemo3} & 0.092$\pm$0.007 \\
  $^{100}$Mo & 3034 & 9.63 & $>$0.11~\cite{Nemo3} & 0.0071$\pm$0.0004 \\
  $^{130}$Te & 2528 & 34.17 & $>$0.40~\cite{Te0nu} & 0.69$\pm$0.13 \\
  $^{136}$Xe & 2479 &   8.87 & $>$11~\cite{Xe0nu} & 2.19$\pm$0.06 \\
\end{tabular}
\caption{Q-value, isotopic abundance, 0$\nu$DBD half-life limits and 2$\nu$DBD half-life measurements (from~\cite{2nudbd}) for a selection of double-beta decay candidate isotopes.}
\label{tab:isotopes}
\end{center}
\end{table}

\section{Present status of experimental searches}\label{sec:presentstatus}

The best 0$\nu$DBD half-life limits available at present come from experiments studying $^{130}$Te, $^{76}$Ge and $^{136}$Xe (see table~\ref{tab:isotopes}).
These translate into $\left<m_{\beta\beta}\right>$ upper limits of (270--760)$\,$meV, (150--330)$\,$meV and (61--165)$\,$meV for $^{130}$Te, $^{76}$Ge and $^{136}$Xe respectively.

The $^{130}$Te result was obtained by CUORE-0~\cite{Te0nu} with an array of 52 TeO$_2$ crystals operated as cryogenic calorimeters.
The detector had a total $^{130}$Te mass of 11$\,$kg, an energy resolution of 5$\,$keV FWHM and a background at Q$_{\beta\beta}$ of 0.058$\,$counts/(keV$\cdot$kg$\cdot$y).
CUORE~\cite{cuore} is a larger and more sensitive version of CUORE-0 that will start data taking in 2017.
The mass of $^{130}$Te is 206$\,$kg and the expected background at Q$_{\beta\beta}$ is 0.01$\,$counts/(keV$\cdot$kg$\cdot$y).
If these performance parameters will be met, CUORE will have a $\left<m_{\beta\beta}\right>$ sensitivity in the range (50 -- 130)$\,$meV.

The $^{76}$Ge result was obtained by GERDA~\cite{Ge0nu} using bare germanium detectors enriched in $^{76}$Ge.
The total detector mass is 35.2$\,$kg, with enrichment in $^{76}$Ge going from 7.8\% to 87\%.
The energy resolution is 3$\,$keV in the best detectors, and the background at Q$_{\beta\beta}$ is as low as 0.001$\,$counts/(keV$\cdot$kg$\cdot$y).
The GERDA collaboration plans to continue taking data in the current configuration until an exposure of 100$\,$kg$\cdot$y will be obtained.
The expected half-life sensitivity will be about 2$\cdot$10$^{26}\,$y.

The $^{136}$Xe result was obtained by the KamLAND-Zen~\cite{Xe0nu} collaboration by dissolving xenon in the ultra-pure liquid scintillator of the KamLAND detector.
The total amount of $^{136}$Xe was of 340$\,$kg but the amount of useful isotope mass was sensibly reduced by the fiducial volume cut imposed during data analysis.
The poor energy resolution, 270$\,$keV FWHM, was compensated by the large mass and by the extremely low background of $\sim$160$\,$counts/(ton$\cdot$y) in the region of interest (ROI).
In 2017 the KamLAND-Zen collaboration plans to reach a sensitivity of about 40$\,$meV on $\left<m_{\beta\beta}\right>$ by performing minor upgrades to the current detector configuration.

\section{Sensitivity of future searches}
Present and near-future experimental searches have sensitivities on $\left<m_{\beta\beta}\right>$ not better than (40--50)$\,$meV.
In this section we discuss the general features of future experiments taking as a target a sensitivity on $\left<m_{\beta\beta}\right>$ of about 15$\,$meV, which corresponds to the full exploration of the inverted region of the neutrino mass-hierarchy.
Using nuclear matrix elements from~\cite{Engel}, this translates into half-life sensitivities roughly in the range (10$^{27}$ -- 10$^{28}$)$\,$y.
We begin discussing the formula that give a rough estimate of the sensitivity $S^{0\nu}$ of a given experiment.
It has different expressions, depending on whether or not the background can be considered negligible.
If the background is not negligible, then we have
$$
S^{0\nu}\;\propto\;\eta\,\varepsilon\,\sqrt{\frac{M\,t}{b\,\Delta E}},
$$
where $\eta$ is the isotopic abundance of the DBD isotope, $\varepsilon$ is the detection efficiency, M is the total detector mass, t is the measurement time, b is the background index expressed in counts/(keV$\cdot$kg$\cdot$y) and $\Delta$E is the FWHM energy resolution.
If instead the background is negligible, we have
$$
S^{0\nu}\;\propto\;\eta\,\varepsilon\,M\,t.
$$
This condition is verified when the total number of expected background counts over the whole duration of the experiment is negligible, i.e. when M$\,$t$\,$b$\,\Delta$E$\,<\,$1, and is obviously preferable because the sensitivity scales linearly with the detector mass.
We note that presently GERDA is the only experiment that was able to obtain this condition of zero-background.

For a given 0$\nu$DBD half-life, the number of expected signal counts N$_s$ can be expressed as N$_s$ = $\ln 2\,\varepsilon\,N_{\beta\beta}\,t\,/\,T^{0\nu}_{1/2}$, where $\varepsilon$ is the detection efficiency and N$_{\beta\beta}$ is the number of isotopes under observation.
Even when the zero-background condition is met, observing a number of signal counts of the order of 1 over a data taking period of the order of one year would demand for a number of source isotopes N$_{\beta\beta}\,\sim\,T^{0\nu}_{1/2}/(1\,y)$.
Therefore about (10$^{27}$ -- 10$^{28}$) source isotopes are needed for a sensitivity of $\sim$15$\,$meV on $\left<m_{\beta\beta}\right>$.
This roughly corresponds to one ton of active source mass.

Background from radioactivity represents the major concern for most present experiments, and will be probably the most challenging scientific problem to address in the future.
However, even assuming that this background contribution could be made negligible, there is still a background coming from the high-energy tail of the 2$\nu$DBD spectrum that cannot be eliminated.
The number of background counts from 2$\nu$DBD in an energy window of width $\Delta$E around Q$_{\beta\beta}$ can be expressed as~\cite{DBDReview}
$$
N_{2\nu}~\sim~\frac{1}{T^{2\nu}_{1/2}}\frac{\Delta E^6}{Q^{5}_{\beta\beta}}.
$$
It is therefore preferable to have small $\Delta$E and large Q$_{\beta\beta}$ and T$^{2\nu}_{1/2}$.
Moreover, even when other background sources are made negligible, energy resolution still remains an important parameter for 0$\nu$DBD searches.

Other considerations are related to the choice of the source isotope to be investigated and of the experimental technique to be used.
The two choices are often not disjoint, because certain experimental techniques can only be exploited for some isotopes.
This is true for example for Ge-diodes or Xe-TPCs, which will be discussed in the next section.
Concerning the isotope choice, essentially two factors come into play.
The first is related to Q$_{\beta\beta}$.
Isotopes with large Q$_{\beta\beta}$ are preferable because they allow to obtain lower background levels.
This is not only due to the fact that the background for 2$\nu$DBD is smaller for larger Q$_{\beta\beta}$, but also because the radioactive background is lower at higher energy.
In this regard a clear distinction can be made between isotopes with Q$_{\beta\beta}$ smaller or larger than 2615$\,$keV, which is the energy of the most intense $\gamma$-line from natural radioactivity.
Other important aspects are the isotopic abundance of the isotope under investigation and the cost for its enrichment~\cite{barabash2012}.
Finally, one could wonder if there are isotopes with a larger expected signal rate for a given amount of source mass.
As pointed out in~\cite{Robertson}, this does not turn out to be the case, as all interesting isotopes are almost equivalent in this regard.

\section{Future searches}\label{sec:futureexp}
We now discuss the experimental techniques that will be used in future 0$\nu$DBD searches and present some of the most promising experiments that implement them.

\subsection{Germanium detectors}
Germanium detectors feature superior energy resolution and allow to discriminate very effectively between single-site (electron-like) and multi-site ($\gamma$-like) energy releases.
This experimental technique can only be applied to $^{76}$Ge, that has a not so high Q$_{\beta\beta}$ of 2039$\,$keV.
Nevertheless, thanks to particle discrimination germanium experiments already demonstrated background rates as low as 0.001$\,$counts/(keV$\cdot$kg$\cdot$y).
The mass scalability of this technique can be extended to the ton scale, but probably not much more than that.
Enrichment in $^{76}$Ge is feasible and already used in experiments, however it is somehow expensive if compared to other isotopes.

The GERDA~\cite{Ge0nu} and Majorana~\cite{Majorana} experiments aim at building a ton-scale germanium experiment, joining their scientific and financial efforts.
GERDA was already discussed in section~\ref{sec:presentstatus}, it demonstrated the possibility to obtain an energy resolution of 3$\,$keV FWHM and background as low as 2$\,$counts/(ton$\cdot$y) in the ROI.
The experiment is currently running with a total detector mass of 36$\,$kg.
Majorana is currently more focused on the radioassay of the materials to be used in the construction of a ton-scale detector~\cite{MajoranaNim}.
A background of less than 3.5$\,$counts/(ton$\cdot$y) in the ROI is expected, however the current prototype detectors measured a slightly higher background corresponding to about 23$\,$counts/(ton$\cdot$y).

\subsection{Bolometric detectors}
Bolometers feature an energy resolution ($\sim$5$\,$keV) comparable to that of germanium detectors and can be exploited to investigate several isotopes.
It has been already demonstrated by CUORE that the detector mass can be extended up to the ton scale, but it probably cannot be increased much more than that.
Radioactive background is the main concern for current bolometric experiments, but the situation can be improved by the introduction of active background rejection techniques.
This is what is planned for CUPID~\cite{cupid1,cupid2}, an upgrade of CUORE with particle-identification capabilities aiming at a background of 0.1$\,$counts/(ton$\cdot$y) in the ROI and a $\left<m_{\beta\beta}\right>$ sensitivity in the 15$\,$meV range.
Two possible strategies are being considered.
The first is to use enriched $^{130}$TeO$_2$ crystals and to perform particle discrimination based on the Cerenkov radiation emitted by 0$\nu$DBD electrons and not by background $\alpha$ particles.
The second is to move to an isotope with Q$_{\beta\beta}$ above 2615$\,$keV and use enriched scintillating crystals such as Zn$^{82}$Se or Li$^{100}$MoO$_{4}$. In this case the particle discrimination would be based on the different light yield of electrons and $\alpha$ particles.
Another promising bolometric experiment is AMoRE~\cite{amore}.
It is still at a preliminary stage, but the plan is to build an array of $^{48depl}$Ca$^{100}$MoO$_4$ scintillating bolometers with a mass of 200$\,$kg and a background of about 1$\,$counts/(ton$\cdot$y) in the ROI.
The expected sensitivity on $\left<m_{\beta\beta}\right>$ is of about 15$\,$meV.

\subsection{Xenon time-projection chambers}
Noble gas or liquid time-projection chambers (TPC) are widely employed in rare event searches.
TPCs for 0$\nu$DBD are filled with xenon enriched in $^{136}$Xe.
The radioactive contaminations can be made very low, and
there are R\&D activities aiming at tagging the Ba$^{++}$ daughter nucleus emitted in double-beta decay.
If these R\&D succeed, only the background from 2$\nu$DBD would remain.
Gaseous TPC are filled with $^{enr}$Xe at pressure as high as 10$\,$bar.
In this environment electrons tracks are a few cm long, and background suppression based on event-topology reconstruction can be performed.
In high-pressure gaseous TPCs the mass scalability is possible, but obviously harder than in liquid TPCs.
The energy resolution can be smaller than 1\% FWHM at Q$_{\beta\beta}$, fairly adequate to make 2$\nu$DBD background negligible.
In liquid xenon TPCs it is easier to increase the source mass, but the energy resolution is not optimal.

EXO is a liquid xenon TPC containing 150$\,$kg of $^{enr}$Xe.
The energy resolution at Q$_{\beta\beta}$ is 3.5\% FWHM and the background is of about 10$^{-3}\,$counts/(keV$\cdot$kg$\cdot$y).
A planned extension of EXO, called nEXO~\cite{nexo}, will contain 5$\,$ton of enriched xenon.
Thanks to the better self-shielding of the larger detector volume, and if the energy resolution will be improved to 2.5\% FWHM, nEXO could be able to obtain a background of about 2$\,$counts/(ton$\cdot$y) in the ROI, or even lower if Ba$^{++}$ tagging will be implemented.
The expected sensitivity on $\left<m_{\beta\beta}\right>$ will be (15 -- 25)$\,$meV.
The NEXT collaboration will start operating a 100$\,$kg high pressure xenon TPC in 2018~\cite{next100}.
The expected energy resolution and background at Q$_{\beta\beta}$ are 0.7\% FWHM and less than 10$^{-3}\,$counts/(keV$\cdot$kg$\cdot$y) respectively.
A ton scale experiment would have background of about 10$\,$counts/(ton$\cdot$y) in the ROI, but this can be reduced by improving the energy resolution and the design of the detector.
The PandaX-III project~\cite{pandax} plans to build a 1-ton xenon experiment made of five identical high-pressure xenon TPC of 200$\,$kg each.
The expected energy resolution and background at Q$_{\beta\beta}$ are 1\% FWHM and about 1.5$\,$counts/(ton$\cdot$y), resulting in a $\left<m_{\beta\beta}\right>$ sensitivity of (20 -- 50)$\,$meV. 

\subsection{Tracking detectors}
Tracking detectors are different from other 0$\nu$DBD techniques because here the source isotope is not embedded in the detector.
It is deposited on foils, and the track and energy of the emitted electrons are measured by conventional gas detectors and calorimeters.
This approach allows to investigate any isotope that can undergo double-beta decay.
Energy resolution is quite poor, because part of the electrons energy is released in the source foils themselves and is not measured.
This problem is mitigated by making the source foils very thin, but this imposes severe limits in the mass scalability of this technique.
Nevertheless the reconstruction of the electron tracks provides excellent background suppression and allows to study the angular distribution of the decay~\cite{SuperNemo}.
This feature would be very important to study decay mechanism alternative to the exchange of a light Majorana neutrino.

The SuperNEMO collaboration plans to build a tracking experiment with a source mass of about 100$\,$kg of $^{82}$Se.
This isotope has a Q$_{\beta\beta}$ above 2615$\,$keV and the 2$\nu$DBD half-life is long enough to make its contribution to the background negligible.
The projected sensitivity on $\left<m_{\beta\beta}\right>$ is in the (50 -- 100)$\,$meV range.
The detector will be composed by 20 identical modules.
The commissioning of the first prototype module was completed recently, it contains a source mass of 7$\,$kg in $^{82}$Se, and the measured energy resolution is 8\% FWHM at 1$\,$MeV.

\subsection{Loaded liquid scintillators}
Large volume liquid scintillator detectors can be loaded with double-beta decay isotopes, and can obtain source masses that are hardly achievable with other techniques.
Future experiments are considering $^{130}$Te and $^{136}$Xe as isotopes, which are relatively cheap to enrich.
Energy resolution is far from being optimal, but is compensated by the very low background and the large mass.
As already discussed in section~\ref{sec:presentstatus}, a very effective demonstration of this technique is KamLAND-Zen.
A planned upgrade of this experiment will see the deployment of one ton of enriched xenon, resulting in an expected sensitivity of about 20$\,$meV on $\left<m_{\beta\beta}\right>$.
This will be possible only after a major detector upgrade that will reduce the energy resolution at Q$_{\beta\beta}$ to 5\% FWHM and therefore mitigate the background from 0$\nu$DBD decay.
A similar approach is being pursued by the SNO+ experiment~\cite{sno}.
In this case the plan is to load 780$\,$ton of liquid scintillator with 0.5\% in mass of natural tellurium.
This will correspond to about 1.3$\,$ton of $^{130}$Te, or about 0.25$\,$ton after fiducial volume cuts.
The data taking is expected to start in about two years, and after five years of data taking, the expected sensitivity on $\left<m_{\beta\beta}\right>$ is of (40 -- 90)$\,$meV.
Using $^{130}$Te instead of natural tellurium, or increasing its concentration, could increase the sensitivity in the future.

\section{Conclusion}
We discussed the target sensitivity of future neutrinoless double-beta decay experiments aiming at the full exploration of the inverted region of the neutrino mass-hierarchy, and the scientific challenges they will have to face.
The targets to be achieved are a source mass of the order of one ton and a background of about one count/(ton$\cdot$y) in the region of interest. 
No experiment at present can fulfill these requirements, but promising proposals could meet them in the next decade.

\end{document}